\documentclass[a4paper,conference]{IEEEtran}
\usepackage{amsfonts}

\IEEEsettopmargin{t}{30mm}
\IEEEquantizetextheight{c}
\IEEEsettextwidth{14mm}{14mm}
\IEEEsetsidemargin{c}{0mm}

\usepackage[utf8]{inputenc}
\usepackage[T1]{fontenc}
\usepackage{url}
\usepackage{ifthen}
\usepackage{cite}
\usepackage[cmex10]{amsmath} 
\usepackage{amsthm}
\usepackage{url}
\usepackage[square]{natbib}
\usepackage{graphicx}
\begin{document}
\title{Time-series Random Process Complexity Ranking Using a Bound on Conditional Differential Entropy}
\author{%
  \IEEEauthorblockN{Jacob Ayers and Richard Hahnloser}
  \IEEEauthorblockA{Institute of Neuroinformatics (INI)\\
              D-ITET, ETH Zürich\\
              CH-8057 Zürich, Switzerland\\
              Correspondence Email: jayers@ethz.ch}
   \and 
  \IEEEauthorblockN{Julia Ulrich, Lothar Sebastian Krapp, Remo Nitschke,\\ Sabine Stoll, and Balthasar Bickel}
  \IEEEauthorblockA{Institute for the Interdisciplinary Study of Language Evolution (ISLE)\\
              University of Zurich\\
              CH-8050 Zürich, Switzerland}
     \and

  \IEEEauthorblockN{Reinhard Furrer}
  \IEEEauthorblockA{Department of Mathematical Modeling and Machine Learning (DM$^3$L)\\
              University of Zurich\\
              CH-8057 Zürich, Switzerland}
}

\maketitle

\begin{abstract}
Conditional differential entropy provides an intuitive measure for relatively ranking time-series complexity by quantifying uncertainty in future observations given past context. However, its direct computation for high-dimensional processes from unknown distributions is often intractable. This paper builds on the information theoretic prediction error bounds established by Fang et al. \cite{fang2019generic}, which demonstrate that the conditional differential entropy \textbf{$h(X_k \mid X_{k-1},...,X_{k-m})$} is upper bounded by a function of the determinant of the covariance matrix of next-step prediction errors for any next step prediction model. We add to this theoretical framework by further increasing this bound by leveraging Hadamard's inequality and the positive semi-definite property of covariance matrices. 

To see if these bounds can be used to rank the complexity of time series, we conducted two synthetic experiments: (1) controlled linear autoregressive processes with additive Gaussian noise, where we compare ordinary least squares prediction error entropy proxies to the true entropies of various additive noises, and (2) a complexity ranking task of bio-inspired synthetic audio data with unknown entropy, where neural network prediction errors are used to recover the known complexity ordering.

This framework provides a computationally tractable method for time-series complexity ranking using prediction errors from next-step prediction models, that maintains a theoretical foundation in information theory.
\end{abstract}

\section{Introduction}
Time-series complexity ranking is a fundamental challenge in diverse scientific domains, from econometric analysis of market stability to neuroscience studies of brain connectivity. Tang et al.\ \cite{TANG2015117} provides a comprehensive review showing that complexity measurements can be broadly categorized into three groups: fractality (analyzing self-similarity and long-term persistence), methods derived from nonlinear dynamics (investigating attractor properties in phase space), and entropy measures (quantifying disorder states).

Among entropy-based measures, conditional differential entropy offers an intuitive approach to quantifying the complexity of a time-series random process. The quantity \( h(X_k \mid X_{k-1}, \ldots, X_{k-m}) \) captures the uncertainty in the next observation given its past, making it a direct measure of the extent to which structure or predictability is present in the data. Lower conditional entropy implies stronger temporal regularities and simpler dynamics, while higher values indicate greater randomness or complexity. Unlike static entropy measures, conditional entropy naturally accounts for temporal dependencies, making it particularly suitable for comparing the complexity of different time series. We will define all the variables of \( h(X_k \mid X_{k-1}, \ldots, X_{k-m}) \) rigorously in Section II. 
\setlength{\parskip}{0pt}
\setlength{\parindent}{1em}

However, practical application of conditional entropy faces significant challenges. Even for well-established statistical models such as Gaussian mixture models, the derivation of closed-form probability density functions remains intractable and requires approximation methods~\cite{gmm_entr}. Recent theoretical advances by Fang et al. \cite{fang2019generic} established that the conditional differential entropy of a time series is upper bounded by a function of the determinant of the covariance matrix of prediction errors for any causal predictor. This connection suggests that prediction error analysis can serve as a tractable proxy for conditional entropy estimation, particularly relevant given rapid advances in machine learning that continuously improve predictor performance. 
\setlength{\parskip}{0pt}
\setlength{\parindent}{1em}

In this paper, we build on the inequality established by Fang et al. \cite{fang2019generic}. First, we verify that the limits hold for synthetic data generated through linear autoregression, where the conditional entropy is known. We then demonstrate that neural network (NN) prediction errors preserve the complexity ranking of synthetic bioacoustic-inspired audio data; where no well-defined probability density function (PDF) exists to directly compute the conditional differential entropy.

\section{Generic Variance Bounds on Prediction Errors}
\subsection{Theoretical Bounds}
Let $X_k$ with $k \in \mathbb{Z}$ be a time series random process in $\mathbf{R}^d$, and consider predicting $X_k$ given context $Y = \{{X_{k-1}, \ldots, X_{k-m}}\}$. Let $\Sigma_\varepsilon$ denote the covariance matrix of the prediction error $\varepsilon_k = X_k - f(Y)$ of a predictor function $f(Y)$. 
By entropy translation invariance \cite[Theorem 8.6.3]{cover2006elements}, for any predictor $f$ dependent on Y:
\begin{equation}
h(X_k \mid Y) = h(X_k - f(Y) \mid Y)
\end{equation}
Since mutual information is non-negative and using the entropy decomposition $h(X_k - f(Y)) = h(X_k - f(Y) \mid Y) + I(X_k - f(Y); Y)$, \cite[Theorem 8.6.1]{cover2006elements} we obtain:
\begin{equation}
h(X_k \mid Y) \leq h(X_k - f(Y))
\end{equation}
For the prediction error $\varepsilon$ with zero mean (from any predictor that minimizes the the mean squared prediction error \cite{rosa2010elements}), the fact that the Gaussian maximizes the entropy over all distributions with the same covariance matrix, implies the lower bound \cite[Theorem 8.6.5]{cover2006elements}. The upper bound follows from Hadamard's inequality applied to the positive semi-definite error covariance matrix $\Sigma_\varepsilon$ \cite[Theorem 17.9.2]{cover2006elements}.

\begin{align}
h(X_k \mid Y) &\leq \frac{d}{2} \ln(2\pi e) + \frac{1}{2} \ln(|\Sigma_\varepsilon|) \nonumber \\
        &\leq \frac{d}{2} \ln(2\pi e) + \frac{1}{2} \sum_{i=1}^{d} \ln((\Sigma_\varepsilon)_{ii})
\end{align}
These bounds establish a fundamental connection between prediction accuracy and information content: systems with higher conditional entropy are inherently less predictable, manifesting as larger prediction errors.
\subsection{Practical Estimation}
In practice, for a time-series random process of interest, we define $y^j = \{x_{k-1}^j, \ldots, x_{k-m}^j\}$ as the context window for the $j$-th observation. We then collect $N$ data points of the form $(y^j, x_k^j)$, where $x_k^j$ is the target value. We partition our dataset into training and test sets, where the predictor $f$ is trained exclusively on the training data to minimize mean squared error. The trained predictor performs inference on the test data context window to estimate the next sample, we compute the sample covariance matrix of prediction residuals:
\begin{equation}
\hat{\Sigma}_\varepsilon = \frac{1}{N_\text{test}-1} \sum_{l=1}^{N_{\text{test}}} (x_k^l - f(y^l))(x_k^l - f(y^l))^T
\end{equation}
where each term represents a prediction error residual for the $l$-th observation from the test set.
By substituting the estimated covariance matrix into the middle term of our inequality, we obtain what we call the Prediction Error Conditional Entropy Proxy (PECEP):
\begin{equation}
\text{PECEP} = \frac{d}{2} \ln(2\pi e) + \frac{1}{2} \ln(|\hat{\Sigma}_\varepsilon|)
\end{equation}
Since the Gaussian distribution maximizes differential entropy for any given covariance matrix, PECEP represents a worst-case scenario entropy bound---assuming prediction errors follow the most entropic distribution possible given their second-order statistics. This framework enables assessment of the ``gaussianizing whitening'' \cite{fang2019generic} criterion. When the difference between the Hadamard upper bound with $\hat{\Sigma}_\varepsilon$ and PECEP becomes small -- 
\begin{equation}
\left[\frac{d}{2} \ln(2\pi e) + \frac{1}{2} \sum_{i=1}^{d} \ln((\hat{\Sigma}_\varepsilon)_{ii})\right] - \text{PECEP}  \approx 0
\end{equation}

The predictor has effectively captured the structural dependencies between context and next-step prediction. This convergence occurs when the off-diagonal elements of $\hat{\Sigma}_\varepsilon$ approach zero, indicating that the prediction errors between dimensions are nearly uncorrelated and no more information can be gained about the next step from the preceding context.

As the prediction model $f$ improves by capturing more temporal dependencies, PECEP approaches the maximal conditional differential entropy lower bound for a covariance matrix due to the Gaussian assumption, providing an indirect but tractable method for estimating the complexity of a time-series random process with an unknown probability density function.

\section{Synthetic Experiments}
\subsection{Verification of Variance Bounds with Known Differential Entropy}
To validate our theoretical framework, we conducted controlled experiments using synthetic time-series data generated from known autoregressive (AR) models. This allows us to compare our PECEP measure against the true conditional differential entropy lower bound, which can be computed analytically for Gaussian AR processes.

\subsubsection{Synthetic Data Generation}

We generate synthetic data using a $p$-th order vector autoregressive (VAR) model: \begin{equation} X_k = \mathbf{A}_1 X_{k-1} + \mathbf{A}_2 X_{k-2} + \cdots + \mathbf{A}_p X_{k-p} + \varepsilon_k \end{equation}

where $X_k \in \mathbf{R}^d$ represents the $k$-th time frame,$\{ \mathbf{A}_i \}_{i=1}^p$ are regression coefficient matrices in $\mathbf{R}^{d \times d}$, and $\boldsymbol{\varepsilon}_k \sim \mathcal{N}(\mathbf{0}, \boldsymbol{\Sigma})$ is additive Gaussian noise with zero mean and covariance matrix $\boldsymbol{\Sigma}$.

For our experiments, we set the dimensionality $d = 32$ and autoregressive order $p = 8$. The coefficient matrices $\mathbf{A}_i$ are constructed with a band-diagonal structure where non-zero elements are randomly sampled from a uniform distribution $\mathcal{U}(-1, 1)$. Each matrix is normalized by its Frobenius norm and then scaled by an exponential decay factor $0.85^i$ to model realistic temporal dependencies where recent lags have a stronger influence than distant ones. This coefficient generation procedure produced stable autoregressive series, avoiding diverging cases, in our simulation setting. 

For robust evaluation, we generated 30 independent realizations of coefficient matrices. For each realization, we autoregressively generated datasets of size $N = 10^6$ under four noise variance levels: $\sigma^2 \in \{10^{-3}, 10^{-2}, 10^{-1}, 1\}$, where additive Gaussian noise $\boldsymbol{\varepsilon}_k \sim \mathcal{N}(\mathbf{0}, \sigma^2 \mathbf{I})$ is applied at each time step. An example of the generated data can be seen in Figure \ref{fig:exp1-vis-1}.

\subsubsection{Prediction Models}

We evaluate two parametric models for estimating the AR coefficients:

\begin{itemize}
    \item \textbf{Oracle:} A baseline with access to the true generating parameters $\{ \mathbf{A}_i \}_{i=1}^p$.
    
    \item \textbf{Ordinary Least Squares (OLS):} Estimates $\{ \hat{\mathbf{A}}_i \}$ by minimizing the squared reconstruction error $\| \mathbf{X} - \mathbf{Y} \hat{\mathbf{A}} \|^2$ computed using numerically stable Singular Value Decomposition (SVD) methods that avoid direct Moore-Penrose pseudoinverse computation \cite[5.5.2-5.5.4]{golub2013matrix} \cite{moore1920reciprocal, penrose1955generalized}.

\end{itemize}

\begin{figure}[ht]
    \centering
    \includegraphics[width=\linewidth]{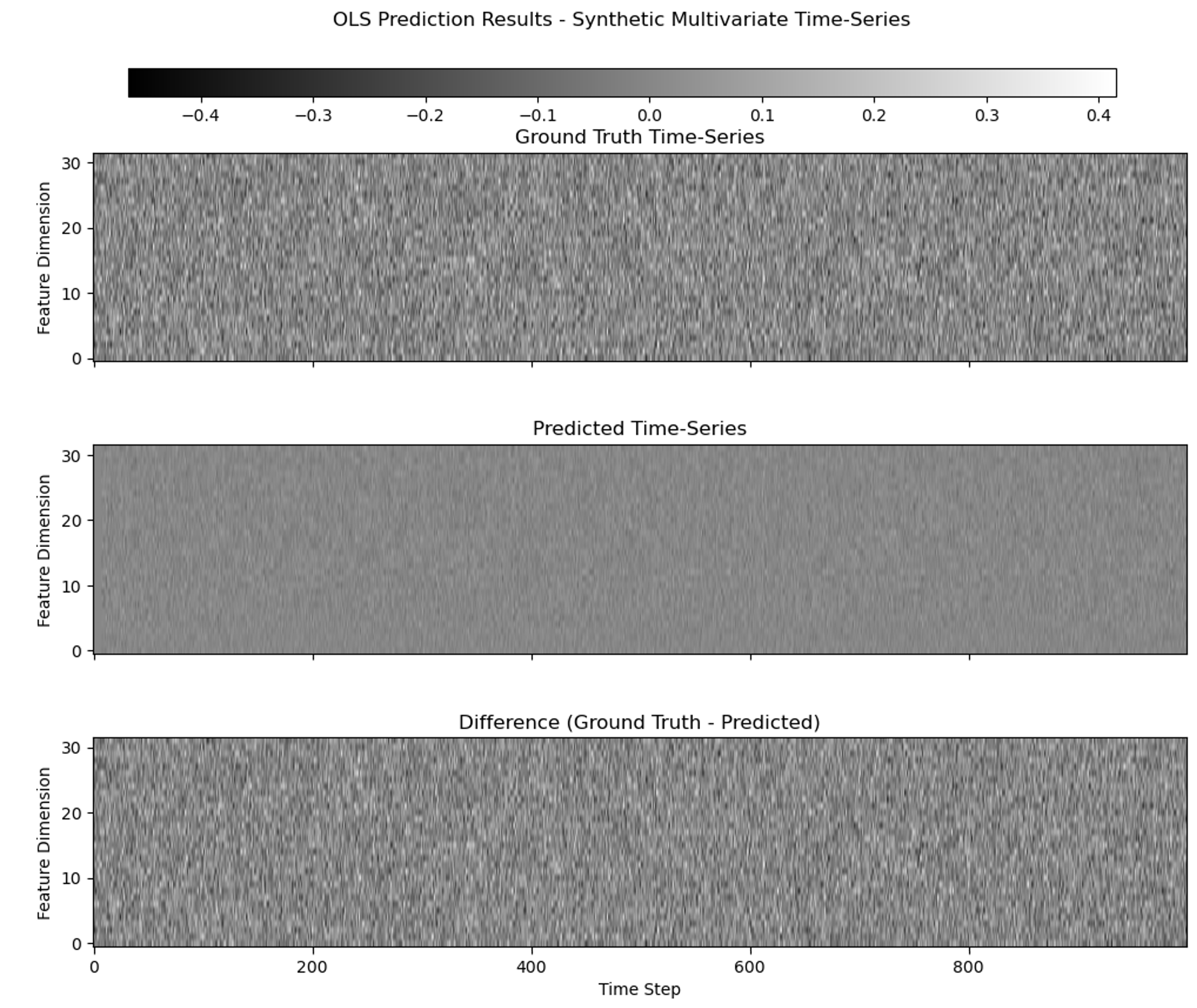}
    \caption{
       Experiment 1 Setup. The top panel shows the ground truth signal across 32 feature dimensions over time. The middle panel shows the time-aligned predictions made by Ordinary Least Squares (OLS). The bottom panel shows the residual errors (ground truth minus OLS prediction) used for calculating the Prediction Error Conditional Entropy Proxy (PECEP).
    }
    \label{fig:exp1-vis-1}
\end{figure}

OLS is trained and evaluated across multiple dataset sizes 
$N \in \{10^3, 5 \times 10^3, 10^4, 5 \times 10^4, 10^5, 5 \times 10^5, 10^6\}$ 
using an 80-20 split, where 
$N_{\text{train}} = \lfloor 0.8N \rfloor$ and 
$N_{\text{test}} = N - N_{\text{train}}$. 
This setup allows us to analyze how close PECEP approaches the theoretical additive noise bound as the amount of training data increases. 
An example of predictions from OLS and the residual errors used to compute PECEP can be seen 
in Figure~\ref{fig:exp1-vis-1}.

\subsection{Verification of Time Series Complexity Ranking with Unknown Differential Entropy}
To evaluate our framework's ability to rank time-series complexity when the underlying probability density function and conditional differential entropy are unknown, we synthesized artificial vocalizations with increasing levels of complexity. This approach enables us to create a ground-truth complexity ordering without requiring analytical computation of the conditional differential entropy.

\subsubsection{Synthetic Data Generation}
We created ten artificial ``species'' of acoustic time series, each defined by progressively more complex spectral, temporal, and structural parameters. The design follows bioacoustic findings linking vocal complexity to communicative sophistication across vertebrate taxa \cite{kavanagh2021dominance, takahashi2016early, stoeger2016information}. Calls are organized hierarchically into bouts composed of multiple calls separated by inter-call intervals, with rest periods between bouts. Each species exhibits systematically increasing complexity: Species 0 has narrow frequency ranges, regular timing patterns, and consistent call power, while Species 9 has wide frequency ranges, highly irregular timing, and inconsistent call power.

Each call begins with spectral energy positioned at a fundamental frequency drawn from species-specific ranges, with integer multiples of this fundamental creating harmonic series that exhibit exponential decay patterns.  Spectral continuity between consecutive frames is restricted by Wiener entropy \cite{tchernichovski2000procedure} and Euclidean distance thresholds that progressively relax between species, allowing for increased spectral variation. Random phases are applied to each frequency bin before inverse FFT conversion. Dynamic formant resonances \cite{fant1971acoustic, fitch2025formant} are implemented as species-specific bandpass filters that simulate how animals modulate harmonic content through vocal tract resonances, with formant frequencies scaling linearly with each species' base frequency range. 

Each call receives a smooth amplitude envelope with gradual onset and offset using squared sine and cosine functions to prevent spectral artifacts from abrupt amplitude changes. Spectral tilting is applied to reduce harsh metallic sounding artifacts. Constant additive Gaussian noise is added to all species recordings. Although the synthetic generation process is fully specified with a set of modifiable variables, the resulting signals involve many interacting factors: non-linear spectral filtering, history-dependent frame transitions, multi-scale temporal variability, and complex harmonic interactions. These factors make analytical computation of conditional entropy intractable, providing a known complexity ordering for evaluating complexity ranking performance.

\subsubsection{Prediction Model}
\begin{figure*}[ht]
    \centering
    \includegraphics[width=\linewidth]{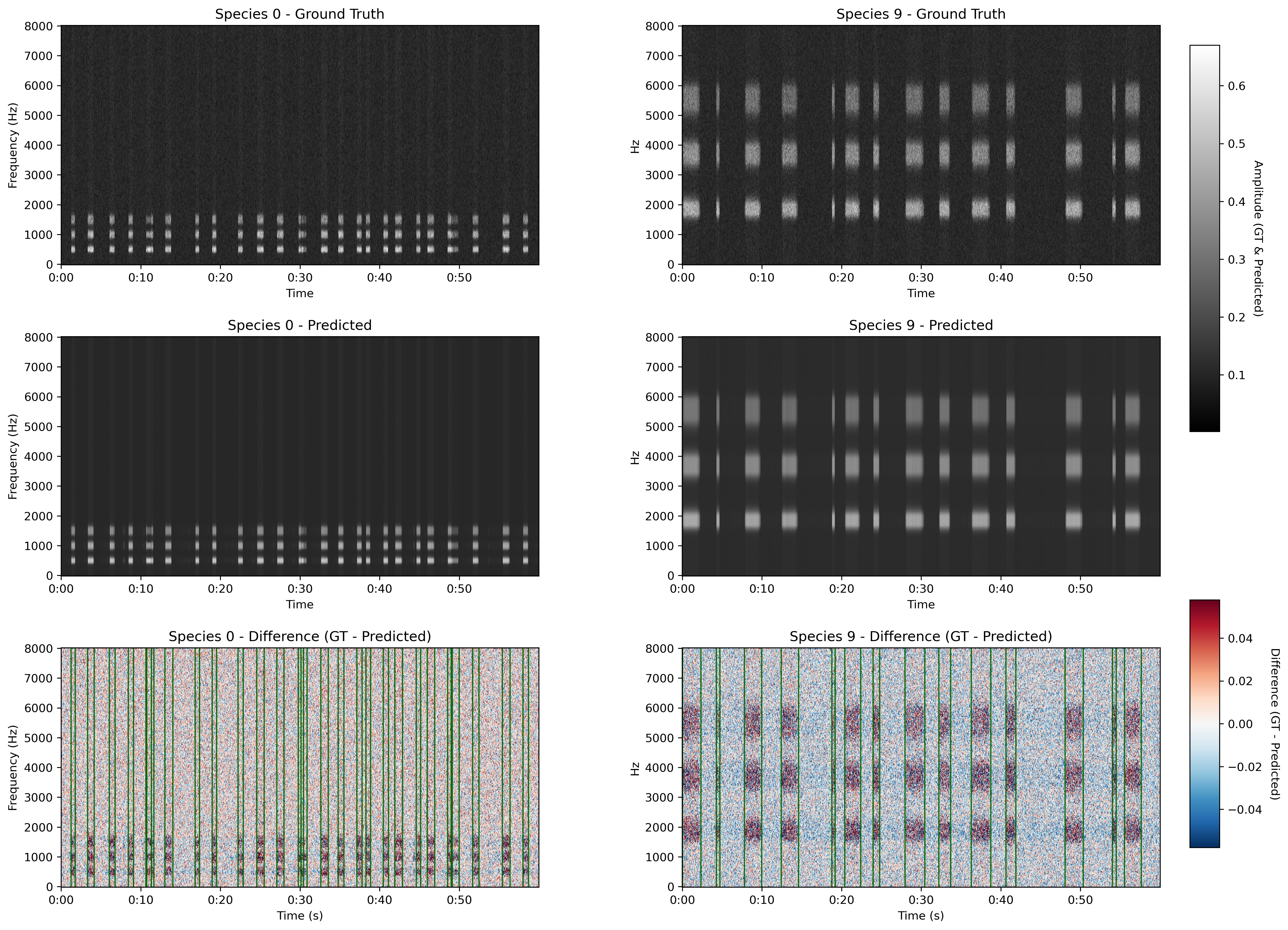}
    \caption{
        Experiment 2 setup. The top row contains example spectrograms of the simplest species 0 and most complex species 9. Species 0 is made up of calls with a smaller frequency range, less variance in power between spectrogram columns containing calls, and less variance between vocalizations. The middle row shows the spectrogram column predictions of a fully connected neural network. The bottom row shows the difference between the ground truth top row and the predicted middle row. The green vertical lines indicate the onset-offsets of the calls as we collect the PECEP scores from the residuals of just those audio segments.
    }
    \label{fig:exp2-vis-1}
\end{figure*}
Before training, audio signals are converted to magnitude spectrograms using a STFT with a Hanning window with a size of 510 samples and 50\% overlap. The resulting spectrograms are converted to the base 10 log scale, then passed through a sigmoid function to map values into a normalized range [0,1].

We construct training sequences following the notation established in Section II. For each spectrogram, we define training examples $(x_k^j, y^j)$ where $x_k^j$ represents the target frame to be predicted and $y^j = \{x_{k-2}, x_{k-3}, \ldots, x_{k-(m+1)}\}$ constitutes the context window of $m = 64$ previous spectrogram columns flattened into a single input vector. The two-step temporal offset ($k-2$ rather than $k-1$) is crucial to avoid the sharing of 
samples in both the spectrogram column to be predicted and the context window used by the predictor. For samples at the beginning of clips, spectrograms are zero-padded to accommodate complete context windows. We perform an 66-14-20 train-validation-test split from the 1200 generated clips.

For our NN model architecture, we use a fully connected network (FCN). The FCN consists of three fully connected layers with ReLU activation functions. The input layer takes the flattened context window and projects it to a first hidden dimension, followed by a second hidden layer, and finally an output layer that reconstructs the predicted spectrogram frame. Dropout regularization of $10\%$ is applied between the first and second hidden layers to prevent overfitting. The model outputs the prediction of the next frame via a sigmoid activation, mapping the output to the same normalized range [0,1] as the input.

Training minimizes the mean squared error (MSE) loss between predicted and actual frames, optimized using AdamW with cosine annealing scheduling over 50 epochs. A model was trained for each synthetic species. An example of the predictions of the FCN model and the difference from the ground truth data can be seen in Figure \ref{fig:exp2-vis-1}.
\setlength{\parskip}{0pt}
\setlength{\parindent}{1em}
To compute meaningful complexity rankings, we collect PECEP scores only during active vocalization periods rather than across entire clips. 

Research on animal communication demonstrates that vocal complexity and temporal patterns are closely linked to social dynamics and communication sophistication. Kavanagh et al.'s \cite{kavanagh2021dominance} comparative analysis across multiple non-human primate species revealed that dominance style fundamentally shapes vocal communication patterns, with tolerant societies showing increased individual vocal rates while despotic societies develop larger hierarchy-related vocal repertoires, while studies of marmosets \cite{takahashi2016early} and elephants \cite{stoeger2016information} reveal sophisticated temporal coordination and turn-taking abilities.

Collecting PECEP scores from predictable near-zero residual error silent periods between vocalizations in our complexity assessment would conflate temporal coordination skills with lower communication complexity. By focusing on active vocalization periods, unpredictable temporal dynamics manifest as increased prediction errors at vocalization onsets, indicating higher complexity, providing a more accurate measure of the underlying communication complexity.

\section{Results}
\begin{figure*}[!t]
    \centering
    \includegraphics[width=0.92\textwidth]{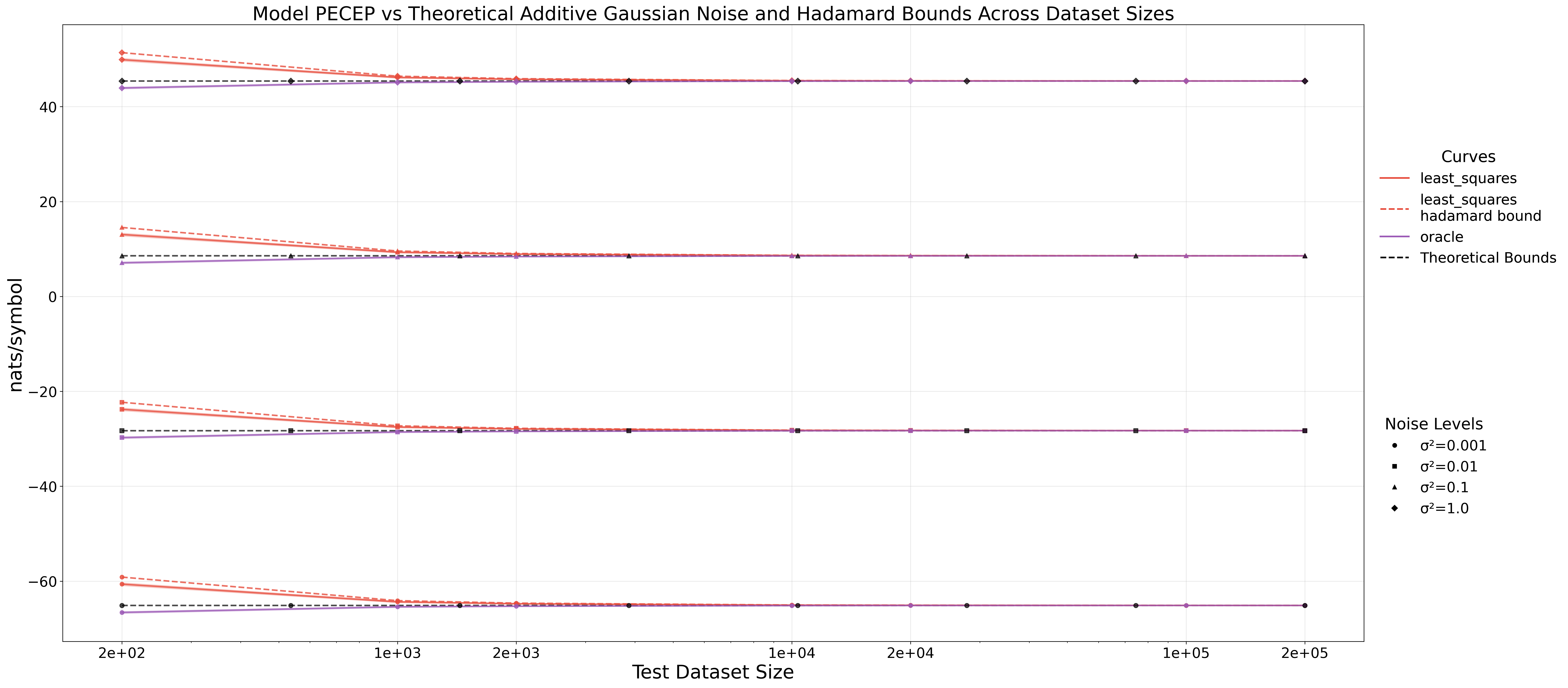}
    \caption{
Experiment 1 Results. PECEP Convergence Validation on Synthetic Autoregressive (AR) Data. Mean PECEP values across 30 trials for the ordinary least-squares AR coefficient model and an Oracle that already knows the ground-truth coefficients at four additive Gaussian noise levels $\mathcal{N}(0,\sigma^2I); \sigma^2 \in \{0.001, 0.01, 0.1, 1.0\}$ and nine dataset sizes. Dashed lines indicate theoretical conditional differential entropy bounds.
    }
    \label{fig:exp1-results}
\end{figure*}

\begin{figure*}[!t]
    \centering
    \includegraphics[width=0.8\textwidth]{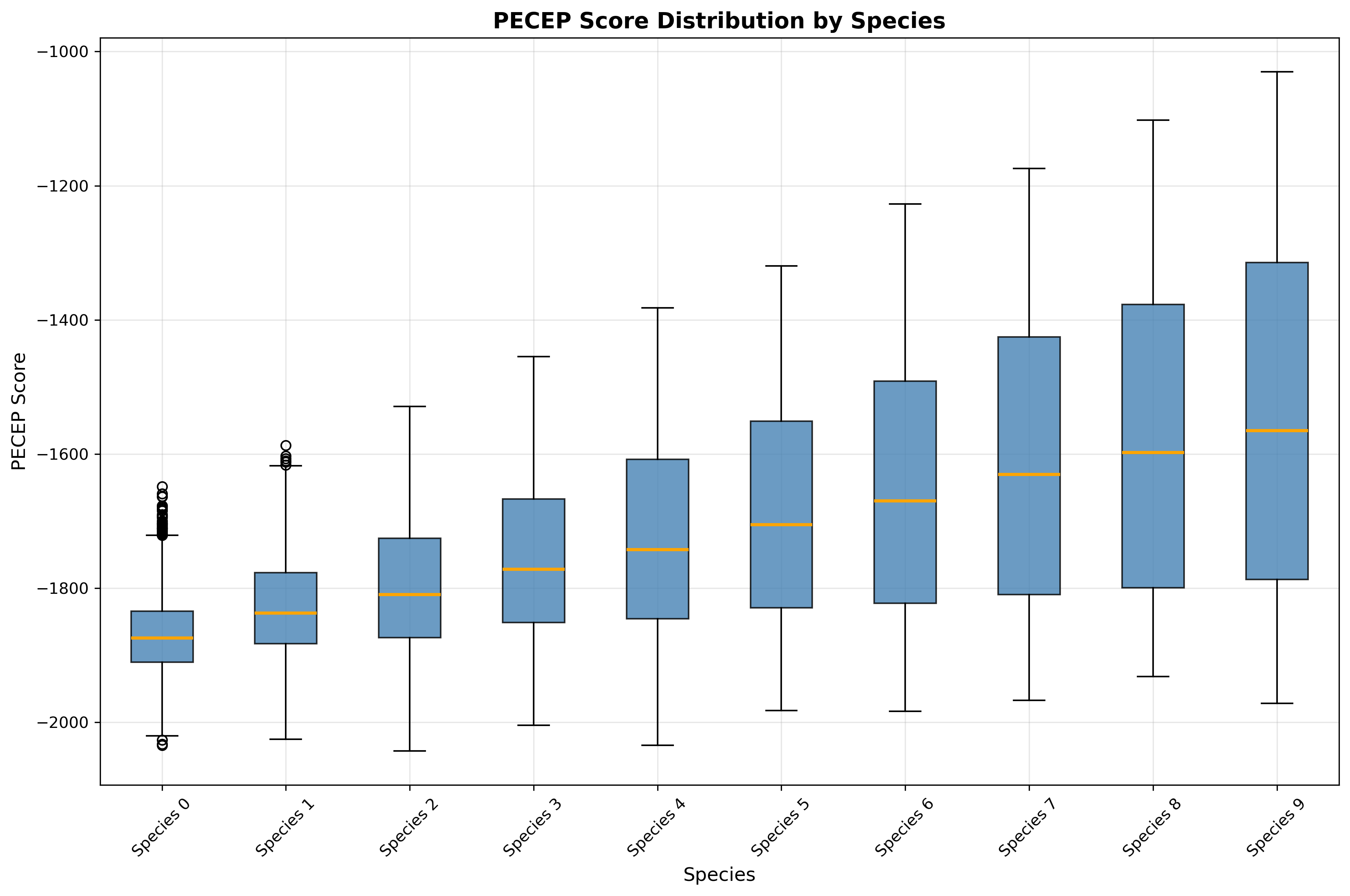}
    \caption{
        Experiment 2 Results. Prediction Error Conditional Entropy (PECEP) boxplots generated from the prediction errors on artificial calls from synthetic species. As the species monotonically grow in acoustic complexity, so do the median PECEP values.
    }
    \label{fig:exp2-results}
\end{figure*}

\subsection{Verification of Variance Bounds with Known Differential Entropy}
Figure \ref{fig:exp1-results} shows the OLS PECEP scores across the 30 trials for increasing additive Gaussian noise variances. We can see that as the dataset size N increases, both OLS and the Oracle PECEP scores converge to the additive Gaussian noise entropy lower bound. 

One unexpected result was that the Oracle consistently underestimated the theoretical entropy bound due to finite-sample bias with only 200 test points. When we increased dimensionality, this underestimation continued to grow—a result documented in this paper's accompanying Github repository \ref{add_mat}. This systematic bias aligns with established findings on finite-sample bias in high-dimensional covariance estimation \cite{ledoit2022power}. Once the sample size increased, the Oracle converged to the true bound.
\subsection{Verification of Time Series Complexity Ranking with Unknown Differential Entropy}
Figure \ref{fig:exp2-results} shows boxplots of the PECEP scores of the calls for each of the synthetic species with unknown entropy but known complexity. From the boxplots, we can see that the median scores monotonically increase with the expected complexity species$_0 <\ $species$_1< \ldots<\ $species$_9$. Showing how the NN prediction errors can be used to rank the complexity of time-series that are challenging to analyze. 
\section{Conclusion}
This work demonstrates that prediction error bounds provide a computationally tractable framework for time series complexity ranking while maintaining theoretical grounding in information theory. We extended Fang et al.'s theoretical variance bounds \cite{fang2019generic} by using Hadamard's inequality to extend the upper bound.

Our bio-inspired synthetic audio experiment demonstrates how NN prediction errors can be used to recover known complexity rankings even when analytical entropy computation is intractable. However, given the monotonic increase in the standard deviation of the PECEP score with increasing complexity of a species in Figure 4, in applications to real data, complex vocalizations will likely require larger datasets for reliable estimation of complexity via PECEP score.

These findings provide an information-theoretic framework for assessing time-series complexity using next-step prediction error. The applicability of the framework to any predictor model makes it promising as more sophisticated forecasting models are used to analyze real-world time series where underlying probability distributions are unknown. Possible avenues for future work are to probe the suitability of the Gaussian assumption of modeled spectrogram data from actual animal recordings and in the non-Gaussian case to explore the tightness of the PECEP bound on conditional differential entropy.

\section*{Additional Materials}
\label{add_mat}
All of the code for this paper, as well as an extended proof from Section II. can be found in this Github repository: \url{https://github.com/JacobGlennAyers/IZS_2026}

\section*{Acknowledgments}
We thank Pau Aceituno of the Institute of Neuroinformatics for his help in reviewing the Fang bounds.

\bibliographystyle{plainnat}
\bibliography{references}

\end{document}